\documentclass[12pt]{article}

\usepackage[left=2cm,right=2cm,top=2cm,bottom=2cm]{geometry}

\usepackage[numbers,sort&compress]{natbib}

\usepackage[centertags]{amsmath}
\usepackage{amsfonts}
\usepackage{amssymb}
\usepackage{verbatim}
\usepackage{amsthm}
\usepackage{newlfont}
\usepackage{graphicx}
\usepackage{times}
\usepackage{color}
\usepackage{empheq}

\usepackage{hyperref}

\newcommand{\be}{\begin{equation}}
\newcommand{\ee}{\end{equation}}
\newcommand{\bea}{\begin{eqnarray}}
\newcommand{\eea}{\end{eqnarray}}
\def\d{\partial}\def\o{\omega}\def\lam{\lambda}\def\v{\vec}
\newcommand{\la}{\langle}
\newcommand{\ra}{\rangle}
\def\mL{\mathcal{L}}
\def\eps{\epsilon}
\def\mI{\mathcal{I}}
\def\sgn{\text{sgn}}
\def\mN{\mathcal{N}}

\newcommand{\nn}{\nonumber}
\definecolor{nv}{rgb}{0, 0.6, 0.2}

\newcommand*\subtxt[1]{_{\textnormal{#1}}}
\DeclareRobustCommand\_{\ifmmode\expandafter\subtxt\else\textunderscore\fi}

\begin{document}
\title{Universal regimes of strong turbulence in the multi-component Gross-Pitaevskii model}
\author{
Vladimir Rosenhaus$^{1}$, Natalia Vladimirova$^{2}$, and 
Gregory Falkovich$^{3^*}$
}
\date{
\small
$^{1}$Initiative for Theoretical Sciences, The CUNY Graduate Center, New York, NY, 10016, USA\\
$^{2}$ Department of Physics \& Astronomy,  Johns Hopkins University, Baltimore, MD, 21210, USA\\
$^{3}$Physics of Complex Systems, Weizmann Institute of Science,  Rehovot 76100, Israel\\[6pt]
%$^{*}$Correspondence: \texttt{gregory.falkovich@weizmann.ac.il}
}

\maketitle
%\date{\today}

\begin{abstract}
The Gross-Pitaevskii (GP) model, also known as the nonlinear Schr\"odinger equation, is arguably the most universal model in classical and quantum physics, describing spectrally narrow or long-wavelength distributions of  interacting waves or particles. Modern applications  ---  from oceanic and atmospheric flows to photonics and cold atoms --- predominantly involve states that are far from equilibrium, culminating in the regime of fully developed turbulence. To date, a consistent theoretical description of such states has only existed for weakly interacting quasiparticles. Here we present a theory of strong turbulence in the two-dimensional $N$-component Gross-Pitaevskii model for both repulsive and attractive interactions, corresponding to the defocusing and focusing cases, respectively. In the focusing case, we show that attraction is enhanced by multi-wave effects, leading to a critical-balance state independent of the pumping level. In the defocusing case, repulsion  is suppressed by collective effects, giving rise to another type of universality in  strong turbulence --- independence from the bare coupling constant.
The theory is confirmed by analytical results in the many-component limit and by direct numerical simulations of the single-component GP model.

\end{abstract}

\noindent
%Compiled \today

We consider an $N$-component Gross-Pitaevskii (nonlinear Schr\"odinger) equation, $i d\vec\Psi/dt=\partial H/\partial {\vec \Psi}^{*}$, where the Hamiltonian describes waves and particles with quadratic dispersion and quartic nonlinearity: 
\be \label{HN}
H=  \int d \mathbf{r}\left(|\nabla \v \Psi|^2+\frac{\lambda}{2N} (\v \Psi^* {\cdot} \v \Psi)^2\right)~.
\ee  
It is $O(N)$-invariant under rotations among the components (which could correspond to different modes, cold-atom species, or spin configurations).
The model arises in many nonlinear systems of waves or particles --- either in the long-wavelength limit or for spectrally narrow distributions. This ubiquity gives it  an ever-expanding range of applications, from classical nonlinear optics, plasma physics, and fluid mechanics to modern photonics and quantum gases.  The coupling constant $\lambda$ is readily tunable in cold-atom and photonic systems, with positive and negative values corresponding to repulsive and attractive interactions, respectively. Far-from-equilibrium states --- known in wave physics as turbulence of envelopes --- are a major focus of current experiments, particularly in clouds of cold atoms \cite{exp1,exp2,exp3,exp4,exp5,exp55,exp6,exp7,exp8,exp9,exp10,exp11,exp12,exp13,exp14,exp15,exp16}. At present, an analytic theory  exists only in the  weak-turbulence limit of a small dimensionless nonlinearity parameter, $\epsilon_k=\lambda \langle|\Psi_k|^2\rangle k^{d-2}$, which represents the ratio of  interaction to dispersion (or potential to kinetic energy for particles), where $\Psi_k$ is the Fourier transform of $\Psi(\mathbf r)$.  In this work, we develop an analytic theory of strong turbulence and verify it through numerical simulations. 
Apart from its broad range of applications, such a theory is of fundamental importance, as it provides a framework for identifying the universality classes of strong turbulence.

Our main object of interest is the occupation numbers $n_k \equiv \frac{1}{N}\la  {\vec\Psi_k}{\cdot}  {\vec \Psi_k}^*\ra$, which satisfy an exact equation:
\be \label{KE1B2}
\frac{\d n_1}{\d t} = -\frac{2 \lam}{N^2 } \frac{1}{(2\pi)^{2d}}\, \int d^dp_2 d^dp_3 d^d p_4\,\delta(\v p_1{+}\v p_2 {-}\v p_3 {-}\v p_4)\text{Im}  \la  {\vec \Psi}_{p_1}{\cdot }  {\vec\Psi}^*_{p_3}\,   {\vec\Psi}_{p_2}{\cdot}  {\vec \Psi}^*_{p_4}\ra~.
\ee
Assuming that the statistics of the waves is close to Gaussian, with a variance  self-consistently taken to be $n_k$, the fourth moment on the right-hand side can be computed perturbatively in either $\epsilon_k\ll 1$  or  $1/N \ll 1$. 
The  equation \eqref{KE1B2}  inherits both conservation laws of the GP equation:  kinetic energy $\sum_kk^2n_k$ and wave action ${\cal N}=\sum_kn_k\equiv \int\frac{d^d k}{(2\pi)^d} n_k$. This means that \eqref{KE1B2} can be written in the form $dn_k/dt=-\partial Q_k/\partial \v k=-k^{-2}\partial P_k/\partial \v k$. Steady-state turbulent cascades of energy and action correspond, respectively, to $k$-independent fluxes $P$ and $Q$. Pumping at wavenumber $k_p$ generates an inverse action cascade and a direct energy cascade,  with the fluxes related by $P\simeq Qk_p^2$. At the lowest order in $\lambda$, the cascades are described by the Kolmogorov-Zakharov weak-turbulence (WT) solutions: $n_k\propto (P/\lambda^2)^{1/3}k^{-d}$ and
 $n_k\propto (Q/\lambda^2)^{1/3}k^{-d+2/3}$ \cite{ZLF}, which are independent of the sign of $\lambda$. 
 
Weak turbulence  in two dimensions is peculiar: the direct cascade scaling $n_k\propto k^{-2}$ coincides with that of  energy equipartition, while the inverse action cascade has the wrong sign of flux. Numerical simulations show that, in fact, pumping heats the system, which stays close to  thermal equilibrium, $n_k(t)=T(t)/[k^2+\mu(t)]$,  with the temperature growing from zero to $Q^{1/3}\lambda^{-2/3}$ and the chemical potential decreasing from $\mu\simeq k_p^2$ to the minimum $ k_0^2$ imposed by a low-$k$ sink. The WT steady state is close to energy equipartition, $n_k\simeq  Tk^{-2}\simeq k^{-2}(Qk_p^2/\lambda^2)^{1/3}$,  with a  distortion by logarithmic factors \cite{ZLF,OT,nonloc}. Since the distortion is  nonlocal, the small local correction proposed in \cite{OT} is not realized \cite{nonloc}. 
As the inverse cascade develops toward lower wavenumber, the nonlinearity $\epsilon_k \simeq \lambda n_k \propto k^{-2}$ increases, marking the transition from weak to strong turbulence. This work describes this transition, which depends crucially on the sign of $\lambda$.

\begin{figure}[t]\begin{center}
   \includegraphics[width=0.95\textwidth]{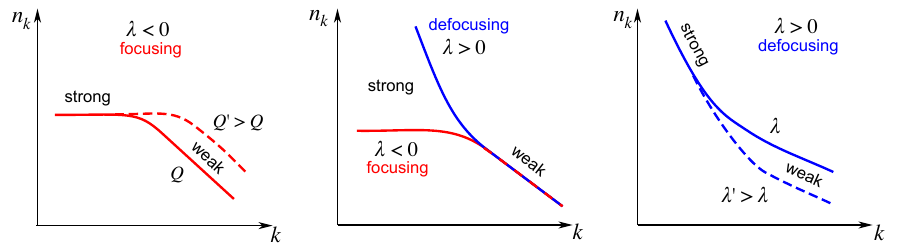}
     \caption{A schematic depiction of the transition from weak turbulence at large $k$ to strong turbulence at small $k$.  The left panel shows the case with a fixed negative $\lambda$, with the dashed line corresponding to larger flux. The right panel shows the positive $\lambda$ case with a fixed flux, with the dashed line corresponding to larger coupling. The central panel shows two cases, with equal flux $Q$ and  couplings $\lambda$ and $-\lambda$, respectively.
                }
    \label{Sketch}
\end{center} \end{figure}

We summarize our main results schematically in Fig.~\ref{Sketch}. The central panel illustrates that the spectra for both signs of $\lambda$ coincide in the weak-turbulence regime, gradually diverging as turbulence gets stronger at lower $k$. While WT spectra depend on both the bare coupling $\lambda$ and the pumping rate $Q$, two universality classes emerge in strong turbulence. The right panel shows the defocusing case, in which the strong-turbulence spectrum is steeper and independent of $\lambda$. The left panel shows the focusing case, in which the strong-turbulence spectrum is flat and independent of $Q$. The strong turbulence spectra can be viewed as  WT spectra, with a renormalized interaction: $\lambda\to \Lambda$. We will show  that repulsion ($\lambda>0$) is suppressed, with the effective interaction going to zero in the long-wavelength limit, $\Lambda\simeq k^2/{\cal N}$, while attraction ($\lambda<0$) is enhanced, becoming unbounded, $\Lambda\simeq \lambda(Q\lambda)^{1/2}k^{-2}$. The resulting strong-turbulence spectra for an inverse cascade in two dimensions are:
\begin{subequations}
\begin{empheq}[left={\displaystyle n_k \simeq \left(\frac{Q}{\Lambda^2}\right)^{1/3} k^{-4/3} \simeq \empheqlbrace}]{align}
&Q^{1/3} \mathcal{N}^{2/3} k^{-8/3} \simeq Q k_0^{-4/3} k^{-8/3}, &\quad \mathrm{for}\ \lambda > 0~, \label{plus} \\
&-\frac{2\pi}{\lambda}~, &\quad \mathrm{for}\ \lambda < 0~, \label{minus}
\end{empheq}
\end{subequations}
where $k_0$ denotes  the lower end of the  $k^{-8/3}$ spectrum resulting from a
sink.

%----------------------
\subsection*{Theory}
%----------------------

Let us now turn to the analytic theory which gives the interaction renormalization \cite{Rosenhaus:2025bgy} and corresponding strong-turbulence spectra.
The essential tool is large $N$: one can compute the  cumulant in  (\ref{KE1B2}) at leading order in $1/N$ to any order in $\lambda$. Schematically, the leading terms in $1/N$ form a geometric series of ``bubble diagrams'', each loop contributing a factor of $\mL$, to be defined below. The resulting sum produces a renormalized kinetic equation (RKE) \cite{Kraichnan, bergesGasenzerScheppach2010,bergesRothkopfSchmidt2008, Walz:2017ffj, RF2},
   \be  \label{largeNKE}  
\!\!\frac{\d n_1}{\d t} = \frac{1}{N}\frac{1}{(2\pi)^2}\!\!\int\! d^2 p_2d^2 p_3 d^2 p_4\, n_1 n_2 n_3 n_4 | \Lambda|^2 \Big( \frac{1}{n_1} {+} \frac{1}{n_2}{-}\frac{1}{n_3} {-} \frac{1}{n_4} \Big)  \delta(\o_1{+}\o_2{-}\o_3{-}\o_4)\delta(\v p_1{+}\v p_2 {-}\v p_3 {-}\v p_4)
\ee
which has the same functional form as the small-$\lam$ kinetic equation, with $\lambda$ simply replaced by  the renormalized vertex  $| \Lambda|^2 = {\lam^2}/{|1 - \mL|^2}$. Define $\v p_- \equiv \v p_4 - \v p_2$ and $\o_- \equiv \o_{p_4} - \o_{p_2}$, we find that for isotropic spectra,
 \bea \nn 
\!\!\!\!\!\!\!\!\!\!\!\!\text{Re } \mL \!\!\!&=&\!\!\!  - \lam \int_{0}^{\frac{|p_-^2{+}\o_-|}{2 p_-}} \!\!\ \frac{dq}{2\pi}\frac{q\,  n_q\,   \text{sgn}(p_-^2{+}\o_-)}{\sqrt{(p_-^2{+}\o_-)^2-(2p_- q)^2}}  - \lam \!\!\int_{0}^{\frac{|p_-^2{-}\o_-|}{2 p_-}}\!\! \frac{dq}{2\pi}\frac{ q\, n_q\,  \text{sgn}(p_-^2{-}\o_-)}{\sqrt{(p_-^2{-}\o_-)^2-(2p_- q)^2}}  \\ \label{A30}
\!\!\!\!\!\!\!\!\!\!\!\!\text{Im } \mL\!\!\!&=&\!\!\!- \lam\int_{\frac{|p_-^2{+}\o_-|}{2 p_-}}^{\infty} \frac{dq}{2\pi} \frac{ q\, n_q}{\sqrt{(2p_- q)^2-(p_-^2{+}\o_-)^2}}+ \lam\int_{\frac{|p_-^2{-}\o_-|}{2 p_-}}^{\infty} \frac{dq}{2\pi} \frac{ q\, n_q}{\sqrt{(2p_- q)^2-(p_-^2{-}\o_-)^2}}~.\, \, \, \, \, \, \, 
 \eea
We will refer to the large-$\lambda$ limit as strong turbulence, even though the deviations from the Gaussian state are of order $1/N$. 

Let us discuss the stationary turbulent solutions of the RKE, starting from the defocusing case, for which we expect interaction suppression and a steeper spectrum.  Going into the strong-turbulence regime with $|\mL|\gg1$,  the renormalized interaction is $| \Lambda|^2 \approx \lam^2/| \mL|^2$, which is independent of the bare coupling $\lambda$. Self-consistently assuming a spectrum steeper than $k^{-2}$, the loop integral $\mL$ diverges at low $k$, so  we may estimate $\mL\simeq \lambda {\cal N}/k^2$ and  $ \Lambda \simeq k^{2} /{\cal N}$. The spectrum \eqref{plus} then follows \cite{Gasenzer}, which we expect to be valid for wavenumbers much greater than the sink, $k\gg k_0$. By way of interpretation,  $\Lambda\, {\cal N}\simeq k^2$ means that the effective total  interaction rate between a wave with a given $k$ and all other waves is the frequency $k^2$.  A vertex that is inversely proportional to the  occupation numbers $n_k$ means that RKE \eqref{largeNKE} is linear with respect to the turbulence level, which gives the linear dependence $n_k\propto Q$ at fixed $k_0$; something never encountered in turbulence before. The strong-turbulence spectrum (\ref{plus}) exhibits the previously advertised universality: it is independent of $\lambda$.

The steep spectrum \eqref{plus} depends explicitly on $k_0$, indicating spectral nonlocality. As a result, this spectrum is far less robust than the usual, local, Kolmogorov-Zakharov spectrum, and is sensitive to the structure of the sink at low $k$, as will be seen in  the DNS below. Without a sink, turbulence has a radically different nature, with a growing condensate and small acoustic perturbations running over it \cite{OT,PT}. The nonlocality of  \eqref{plus} also affects the wavenumber at which turbulence transitions from weak to strong. The RKE shows that the effective dimensionless coupling is not $\epsilon_k\simeq \lambda n_k$, but rather $\mL$, which governs corrections to the WT spectrum. The loop integral $\mL$ is marginally convergent for $n_k\propto k^{-2}$ and diverges at small $k$ for the spectrum \eqref{plus}. This makes the weak-turbulence validity condition, $|\mL|\ll1$, dependent on the extent of the strong-turbulence interval at low $k$, since the   integration in (\ref{A30}) extends over all wavenumbers. For $n_k=Tk^{-2}$, the nonlinearity reaches order unity, $\epsilon_k=\lambda n_k\simeq1$, at $k\simeq\sqrt{\lambda T}$. If the inverse cascade is short and ends at $k_0>\sqrt{\lambda T}$, then it is entirely within the WT domain $\mL\simeq \epsilon_k< 1$ for all $k$. In contrast, a long cascade with $k_0<\sqrt{\lambda T}$ develops a steeper, strong-turbulence spectrum \eqref{plus} at low $k$, which adds a divergent, $k_0$-dependent contribution to the loop integral, $\mL\supset \lambda {\cal N}/k^2$, which becomes of order unity  at $k\simeq \sqrt{\lambda{\cal N}}$. Consequently, for long cascades with ${\cal N}>T$ (equivalently, $k_0^3<\lambda Q/k_p$), weak turbulence breaks down not at $\sqrt{\lambda T}$ but earlier, at the larger wavenumber $\sqrt{\lambda{\cal N}}$, which depends on $k_0$.

Let us now determine the stationary strong-turbulence solution of the RKE \eqref{largeNKE} for the focusing case. Along the cascade toward lower $k$, the interaction must become progressively stronger, as the denominator in $|\Lambda|^2$ decreases with increasing $\mL$. If $\mL$ continued to grow past unity,   the renormalized interaction vertex $\Lambda$ would  begin to decrease, which seems unphysical. Therefore, we assume that the asymptotic state at low $k$ has  $\mL \approx 1$. This state is known as critical balance, in which nonlinearity (potential energy) and dispersion (kinetic energy) exactly balance at all wavenumbers or momenta.
The condition $\mathrm{Re }\,  \mL=1$ yields the stationary spectrum in \eqref{minus}, 
which is independent of the flux $Q$.
Of course, the critical-balance scenario has been proposed many times before on qualitative grounds \cite{Phil,GS,NZ,NS}; the novelty here is that it arises as an actual stationary solution of a kinetic equation.
Formally, equipartition, $n_k=\mathrm{const}$, is a stationary solution for any constant, yet the low-$k$ asymptotics of an inverse turbulent cascade exist only for the specific value \eqref{minus}.
In other words, no other equipartition state can absorb flux, whereas the particular equipartition \eqref{minus} can absorb any amount of flux! 
Since $\mathrm{Im}\,\mL = 0$ for $n_k = \mathrm{const}$ in two dimensions, the leading correction to the low-$k$ asymptotic \eqref{minus} follows from a Taylor expansion of the loop integral in $k^2/k_*^2$, giving
\be
\Lambda \propto (1 - \mL)^{-1} \simeq \lambda (k_*/k)^2\,,
\label{Lambda}
\ee
as stated in \eqref{minus}. 
Here we have assumed that the critical-balance regime begins at $k_* = (\lambda Q)^{1/4}$, where collapses absorb action at a rate $k_*^2 n_k k_*^{d} \simeq k_*^4 / \lambda $ that is comparable to $Q$. 
The scaling \eqref{Lambda} may be interpreted as attraction enhancement of long-wavelength modes  ($k<k_*$) by the collective effects of all the waves in the collapsing cavern, in proportion to the ratio of the kinetic energies.

%------------------------------------
\subsection*{Numerical simulations}
%--------------------------------------
Let us now compare our predictions from the renormalized kinetic
equation (valid for large $N$) with the steady-state spectra obtained
by direct numerical simulations (DNS) of the one-component ($N=1$) GP
model~\cite{Pre,nonloc}, see also \cite{Rus, Naz1, Naz2} for 3D,
\be
i \Psi_t + \nabla^2 \Psi - \lambda |\Psi|^2 \Psi = i \hat{f}_k \Psi + i \hat{g}_k + i\hat{h}_k \Psi~.
\label{DNS}
\ee
The forcing and damping are applied in spectral space. The forcing is additive, $g_k = |g_k|e^{i\phi_k}$
with random phases $\phi_k$ and  amplitudes $|g_k| \propto \sqrt{(k^2 - k^2_p)(k^2_{r} - k^2)}$ scaled
to provide an influx of waves $\dot{\cal{N}}_\text{DNS} = Q$, where
$\cal{N}_\text{DNS} = \int |\Psi(\mathbf{r})|^2 d\mathbf{r} / \int d\mathbf{r}$
is the wave action per area.
  The damping is multiplicative and
applied in two regions of the spectrum, $h_k = -\beta(k/k_h)^4(k/k_h - 1)^2$ at $k>k_h$,
and $f_k = -\gamma k^s$ at $k<k_p$, with $s{=}{-}1$ or $s{=}{-}2$.
Equation~\eqref{DNS} is solved by a fourth-order-accurate in time split-step method.
The computational domain is a square, $L\times L$, with periodic boundary conditions,
so that the lowest wavenumber is $k_{\min} = 2\pi /L$.
The wave action per area can be computed in  spectral space as
$\cal{N}_\text{DNS} = \int d^2k\, n_k / k^2_{\min}$.
The highest wavenumber is the same in all simulations, $k_{\max} = 512$.
The high-$k$ damping starts at  $k_h=256$; unless specified, $\beta=400$.
The pumping range, $k_p = 68$, $k_{r}=84$, is also the same in all runs. Low-$k$ damping (sink)
varies across simulations.

\begin{figure}[h!]\begin{center}
    \includegraphics[width=0.85\textwidth]{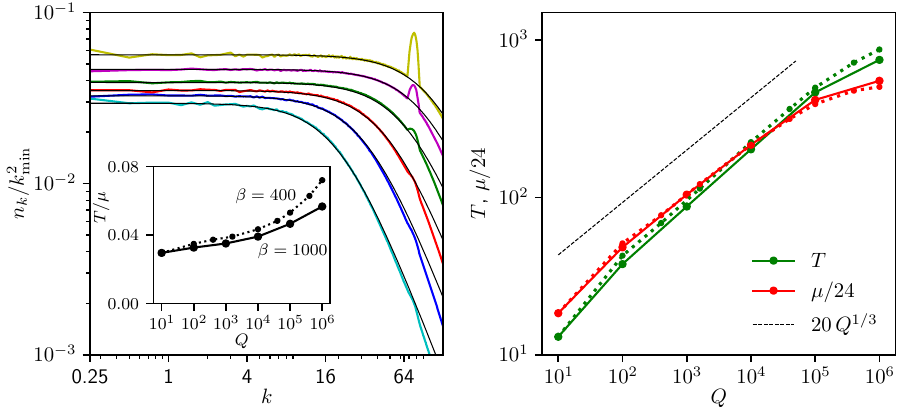}
    \caption{Direct numerical simulation of the focusing Gross-Pitaevskii
      model with one field, $N=1$, $\lambda=-1$, with $k_{\min}=1/4$ and no
      low-$k$ damping, $f_k=0$, for different pumping rates.
      Left panel: Spectra for $\beta=1000$ and pumping rates $Q$ shown in the inset;
      black lines are approximations
      $n_k =T /[\mu +k^2]$ fitted up to pumping peak; inset shows the plateau values. The $n_k/k_{min}^2$ here should be compared with $n_k/(2\pi)^2$ in the theory section: the large $N$ theory prediction (\ref{minus}) gives $n_k/k_{min}^2 = 1/2\pi\approx .16$, larger by about a factor of $4$ than the value observed here for $N=1$. 
       Right panel: Parameters
      $T(Q),\mu(Q)$ for $\beta=1000$ (solid lines) and $\beta=400$ (dashed lines).}
    \label{fig_clps}
\end{center} \end{figure}

For the focusing case, Figure~\ref{fig_clps} shows the stationary spectra for pumping rates spanning five orders of magnitude. The plateau value changes only slightly, supporting our prediction. As the flux increases, the curve in the inset (the plateau amplitude) tends to saturate; the slight increase of the plateau level observed for the largest $Q$ appears to result from a narrow peak of $n_k$ forming around the pumping scale. Increasing the strength of the high-$k$ dissipation, $\beta$, suppresses the pumping peak and flattens the curve.
The weak $Q$-dependence of the equipartition spectra at both $N=1$ and $N\gg1$ suggests that critical balance scaling likely occurs for any number of components in the focusing case. This seems physically natural, since collapses occur and evolve in a similar manner for all $N$.

\begin{figure}[h!]
   \centerline{\includegraphics[width=\textwidth]{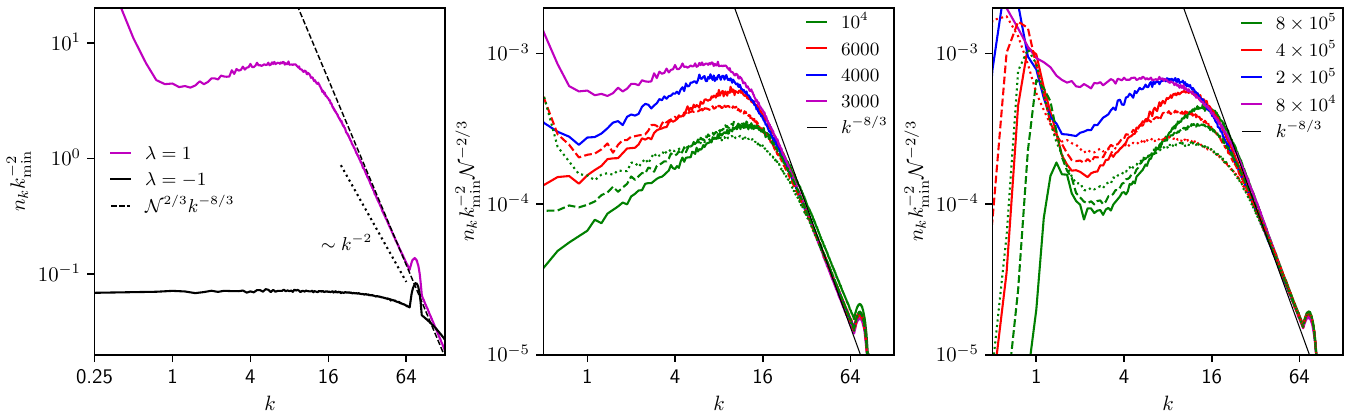}}
   \caption{DNS of GP for $N=1$,  $k_{\min}=1/8$, and the pumping rate is $Q=10^6$. Left
     panel: Lower (black) curve is for the focusing case with no
     low-$k$ sink, $f_k=0$; upper (purple) curve is for the defocusing case with
     the sink $f_k = -3000 k^{-1}$ for $k<k_p$. Middle and right panels: 
     defocusing case for  $f_k = -\gamma k^{-1}$ (middle) and
     $f_k = -\gamma k^{-2}$ (right) at $k<k_p$. The colors denote different values
     of $\gamma$ shown in the legend.  Solid, dashed, and dotted lines
     are for $\lambda=1, 2, 4$, respectively. The solid straight line
     is the formula \eqref{plus} without any fitting parameter.}
    \label{8323}
  % \label{fig_k-1}
\end{figure}

\begin{figure}[h!]
   \centerline{\includegraphics[width=0.75\textwidth]{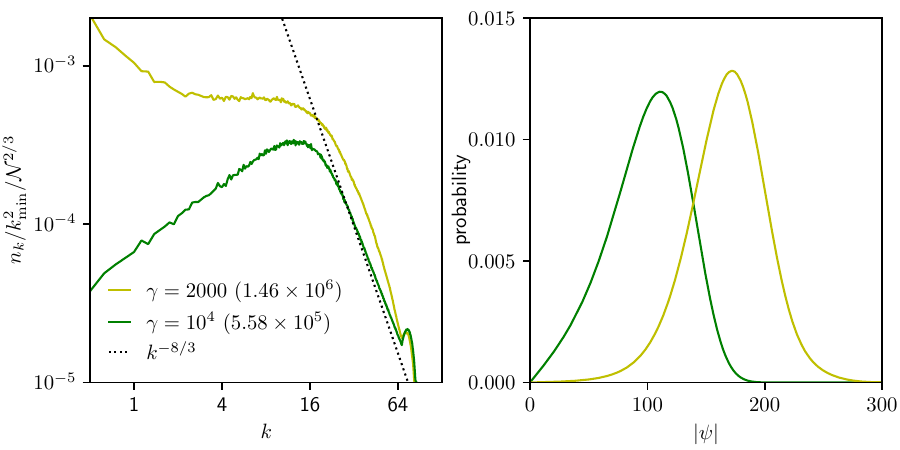}}
   \caption{ Left: Spectra obtained by DNS of the defocusing GP with
     $N=1$, $\lambda=1$, $Q=10^6$, $k_{\min} = 1/8$.
     Damping rate is $f_k = -\gamma k^{-1}$ at $k<k_p$ with two
     different values of $\gamma$. The
     values of action $\cal N$ are given in parentheses.  Only the curve for
     high $\gamma$ agrees with the theoretical formula obtained
     for the $N\gg1$-component GP in \eqref{plus}.  Right: The
     probability density functions of $|\Psi|$ for the two spectra in
     the left panel.  }
\label{8323s}
\end{figure}

The left panel of Figure~\ref{8323} illustrates how dramatically different the strong-turbulence spectra are for the focusing and defocusing cases. In the focusing regime, rare intense fluctuations (collapses) consume both energy and action so efficiently that a much lower mean turbulence level is sufficient to dissipate the same flux.
For the defocusing case, the middle and right panels of Figure~\ref{8323} compare the stationary spectra obtained from DNS of the one-component GP equation with our theoretical prediction~\eqref{plus}. The agreement in the cascade range is remarkable, exhibiting  the correct scaling with $k$ and ${\cal N}$, as well as independence of~$\lambda$. This agreement and universality are predicated on having a strong enough  low-$k$ dissipation rate $\gamma$. The left panel of Figure~\ref{8323s} compares spectra obtained for low and high values of $\gamma$. When $\gamma$ is small, the spectrum deviates from the theoretical prediction at large $k$ and exhibits a pronounced pile-up at low $k$. The probability distribution of the spatial amplitudes $|\Psi|$, shown in the right panel, reveals that, for small $\gamma$, large-amplitude fluctuations occur with substantial probability. These partially coherent regions ---``pre-condensates'' in the terminology of~\cite{Pre} --- generate small vortices and dipoles, thereby reshaping the spectrum across all scales.

%----------------------------------------------------------------------
\subsection*{Discussion}
%---------------------------------------------------------------------

Renormalization of scale-independent interactions into scale-dependent ones is commonplace in quantum field theory: in QED the renormalized interaction weakens at large scales, in QCD it strengthens. The former is understood as a consequence of screening, the latter as a result of a flux string connecting quarks. We have found a renormalized interaction in  turbulence that decreases as the kinetic energy becomes small in the repulsive case, and increases in the attractive case. To uncover the physical mechanism behind this behavior, it is necessary to examine not only the averaged $n_k$ but also the underlying field configurations. The contrast between the defocusing and focusing cases is stark: in the former, the distribution is nearly spatially uniform, while in the latter it is highly inhomogeneous, with localized patches (or caverns) undergoing collapse. We may speculate that in the focusing case, 
low-momentum modes are less likely than high-momentum modes to escape these collapsing regions, and thus experience a stronger effective interaction. Conversely, in the defocusing case, the high-momentum modes traverse a larger range of density fluctuations within the nearly uniform background, allowing them to experience a stronger interaction.

A remarkable achievement of twentieth-century physics was the classification of critical phenomena into universality classes ---
sets of microscopically distinct systems that, nevertheless, share the same large-scale behavior. Can we find universality classes for systems that are far from equilibrium? 

For weakly nonlinear systems the answer is affirmative and somewhat trivial: the weak-turbulent state is independent of the forcing, the dissipation, and any of the details of the interaction aside from its overall scaling.  Here we find that the universality is broken as one goes deeper down the cascade and encounters stronger nonlinearity.  Moreover, an important lesson of this work is that  long cascades  invalidate the common implicit assumption that one may describe weak turbulence locally at high $k$ without accounting for strong turbulence at low $k$.

In the context of strong-turbulent inverse cascades for the multi-component Gross-Pitaevskii model we have found two universality classes: i) for any negative coupling, the state is independent of the pumping strength (the flux), ii) for any positive coupling, the state is independent of the coupling, provided that there is a strong enough sink at low wavenumber. Analytically, we can describe this for a large $N$, yet the universality classes seem to be insensitive to the number of components. 
Gross-Pitaevskii is one of the simplest and most universal models. There are many others, from spin waves to plasma waves;  how their turbulent state falls into these two universality classes, or if there are others, is an important problem which can now start being addressed.

\subsection*{Acknowledgments}

This work used the Expanse cluster at the San Diego Supercomputer Center 
through allocation DMS140028
from the Advanced Cyberinfrastructure Coordination Ecosystem: Services \& Support
(ACCESS) program, which is supported by National Science Foundation grants
\#2138259, \#2138286, \#2138307, \#2137603, and \#2138296. VR is supported by NSF grant \#2209116 and by BSF grant \#2022113. GF thanks NYU and SCGP for hospitality, his work was supported by the Excellence Center at WIS, the Simons grant \#617006, the ISF grant  \#146845, the NSF-BSF grant \#2020765, and the EU Horizon grant  \#873028.

%=============================================================================

%\newpage
%-------------------------
\subsection*{Supplementary material}
%-------------------------
\subsubsection*{Large $N$ kinetic equation (RKE)}
\begin{figure}[h!]
   \centerline{\includegraphics[width=0.5\textwidth]{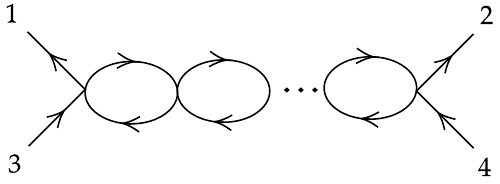}}
   \caption{ For the multi-component GP model  the  process of two wave scattering is dominated by bubble diagrams in the limit of a large number of components. These can be summed, allowing one to study the theory at strong nonlinearity. }
\label{manybubbles}
\end{figure}

Here we provide additional details for the large $N$ (renormalized) kinetic equation (RKE) for the multi-component Gross-Pitaevskii model, given earlier in \eqref{largeNKE}. 
As indicated in the main body, the kinetic equation is expressed in terms of a fourth cumulant, \eqref{KE1B2}.  In the limit of a large number of components (large $N$) the only Feynman diagrams that contribute are bubble diagrams, as shown in Figure~\ref{manybubbles}.
 Each loop gives a factor $\mL$, and performing the geometric sum yields the renormalized interaction vertex $\Lambda = \lam/(1-\mL)$, where $\mL$ is given by \cite{RF2v2}~\footnote{Relative to \cite{RF2v2}, here we have defined $\lambda\rightarrow \lambda/2$, and we have also included factors of $(2\pi)^d$ in the convention for integration: $d \v p \equiv \frac{d^d p}{(2\pi)^2}$. The critical balance solution which comes from setting $\mL = 1$ and was found in \cite{RF2} $n_k =  -1/4\pi \lam$, becomes $n_k = -2\pi/\lam$ quoted in \eqref{minus}, after sending $\lam\rightarrow \lam/2$ and multiplying by $(2\pi)^2$.}
 \be \label{mLm2}
\mL= \mI(\o_-) + \mI^*(-\o_-)~, \ \ \ \ \ \mI(\o_-) \equiv - \lam \int  \frac{d^d q}{(2\pi)^d}\, \frac{n_q }{\o_- {-} q^2 + (\v p_- {-} \v q)^2 +i\eps}~,
\ee
 %\be \label{mLm1}
 %\mL = -2\lam \int d \v q\, n_q\left[\frac{1}{\o_- {-} q^2 + (\v p_- {-} \v q)^2{+}i\eps }-\frac{1}{\o_- {+}q^2 {-} (\v p_- {+ }\v q)^2 {+}i\eps}\right]~,
% \ee
where $\o_- = p_4^2- p_2^2$ and $ \v p_- = \v p_4 - \v p_2$. 
Assuming an isotropic $n_q$ and $d=2$, we may perform the angular integral between $\v p_-$ and $\v q$,
\be
\mI(\o_-) = - \lam \int_{0}^{\frac{|p_-^2{+}\o_-|}{2 p_-}}\!\! \frac{dq}{2\pi}\frac{  q\, n_q\,   \text{sgn}(p_-^2{+}\o_-)}{\sqrt{(p_-^2{+}\o_-)^2-(2p_- q)^2}} 
+    i \lam\int_{\frac{|p_-^2{+}\o_-|}{2 p_-}}^{\infty} \frac{dq}{2\pi} \frac{ q\,n_q}{\sqrt{(2p_- q)^2-(p_-^2{+}\o_-)^2}}~.\label{Loop}
\ee
where we used  the identity, 
 \be
\int_0^{2\pi}d\theta \frac{1}{\cos \theta -a-i\epsilon} =  - \frac{2\pi\ \text{sgn}(a) }{\sqrt{a^2{-}1}}\Theta(|a|>1) +\frac{2 i \pi}{\sqrt{1{-}a^2}}\Theta(|a|<1)~,
 \ee
which can be found by contour integration (one  changes variables to $z=e^{i \theta}$). Here $\sgn(x)$ is $1$ for positive $x$,  $-1$ for negative $x$, and zero for $x=0$.  The result \eqref{A30} immediately follows.

\subsubsection*{Virial Theorem}

A useful insight into the fundamental difference between the focusing and defocusing cases is provided by the Talanov virial theorem \cite{Talanov}, valid in two dimensions and for any number of components $N$:
\be
\frac{d^2}{dt^2}\int d^2r\, r^2|\Psi|^2 = 4H~.
\label{Talanov}
\ee
In the defocusing case, the Hamiltonian $H$ is always positive, so any localized packet inevitably expands. In contrast, in the focusing case, $H$ is  negative for sufficiently  strong nonlinearity, leading to collapse: the packet contracts to zero size in a finite time \cite{ZN}. In two dimensions specifically, each collapse event carries a finite amount of both energy and action to small scales, where they are dissipated \cite{ZK}. This makes collapses an effective physical mechanism to enforce the critical balance state --- collapse occurs whenever attraction overcomes kinetic energy, so that the effective ratio of potential to kinetic energies, $\mL$, never exceeds unity. 

The spectrum \eqref{plus} in the defocusing case  is very different. Direct numerical simulations  show that the Gross-Pitaevskii equation reaches a steady state without any low-$k$ sink in the focusing case, whereas such a sink is required when $\lambda>0$. However, achieving a steady state within the RKE \eqref{largeNKE} requires a low-$k$ sink for both signs of $\lambda$: while the RKE can capture the qualitative evolution of $n_k$, it likely cannot describe the effective damping due to collapses in the focusing case. Since the RKE conserves both ${\cal N}$ and the kinetic energy $\sum_k k^2 n_k$, it necessarily requires two sinks to reach a stationary state, by the standard arguments, see Chapter~2 of \cite{ZLF}.

%--------------------------------------
\subsubsection*{Problems for further studies}
%--------------------------------------
Here we present the results of some numerical simulations which still require a theoretical description.

\begin{figure}[h!]\begin{center}
    \includegraphics[width=0.65\textwidth]{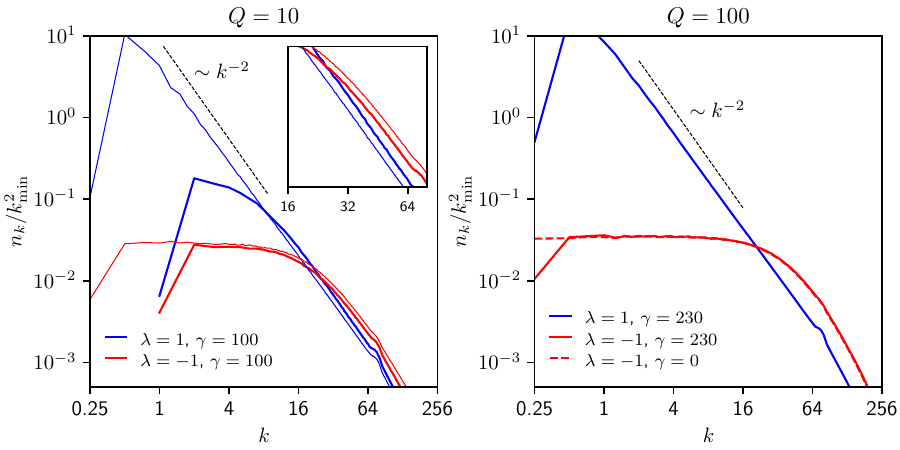}
    \caption{DNS of GP for $N=1$.  Spectra for  the focusing and defocusing cases at the same
      pumping rates.  Blue lines are defocusing and red lines are focusing. In the left panel, thin and thick lines correspond to $k_{\min}=1/4$ and
      $k_{\min}=1$, respectively; in the right panel $k_{\min}=1/4$. 
      The low-$k$ damping is applied to nine modes,
      $f_k=-(1,1,\frac{1}{\sqrt{2}})\gamma$ for $k=(0,1,\sqrt{2})k_{\min}$.}
    \label{fig_pm}
    \label{fig_weak}
\end{center} \end{figure}

Figure~\ref{fig_weak} shows the conditions under which the focusing and defocusing spectra coincide in the weak-turbulence regime.  Identical external pumping and dissipation are applied in both cases: the low-$k$ damping rate is $\gamma$ and the high-$k$ damping rate is $\beta$.  Even for a very weak pumping rate, $Q = 10$, --- when the total interaction energy is fifty times smaller than the kinetic energy --- the spectra coincide only at sufficiently large $k$. 
For $Q\lambda \ll k_p^4$, away from the dissipation regions, all spectra are close to thermal equilibrium, $n_k = T / (\mu + k^2)$. When the pumping is weak enough that turbulence remains weak across all $k$, we find $T \simeq (Q/\lambda^2)^{1/3} k_p^{2/3}$ and $\mu \simeq k_0^2$, for both signs of $\lambda$. 

The right panel of Figure~\ref{fig_pm} show that for flux as low as $Q = 100$ --- when the total interaction energy becomes comparable to the kinetic energy --- the spectra already differ across the full range of $k$. Both the temperature (kinetic energy) and chemical potential are higher in the focusing case, while the total $\mathcal{N}$ and interaction energy remain similar. Although strongly nonlinear effects dominate only at low $k$, they modify the spectrum across all scales. This is particularly striking in the focusing case, where the plateau is insensitive to the magnitude, or even the presence, of low-$k$ dissipation, indicating  that effective dissipation  must be provided by collapses, which overheat the spectrum at large $k$. 

Whether there is an interval of weakly turbulent inverse cascade is determined by the ratio $k_*/k_p\simeq(Q\lambda/k_p^4)^{1/6}$.

\begin{figure}[h!]\begin{center}
    \includegraphics[width=0.65\textwidth]{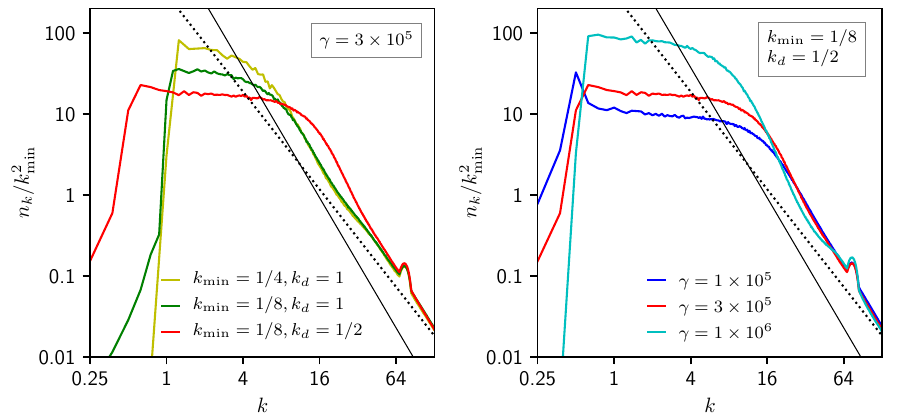}
    \caption{ DNS of GP with $N=1$ in the defocusing case with $Q=10^6$.  Spectra for damping $f_k =
      -\gamma k^{-1}$ applied only to low modes, $k<k_d$.  Different
      $\gamma$, $k_d$ and $k_{\min}$ are shown with colored lines. Left
      panel: yellow: ${\cal N}=1.60 \times 10^6$, green:
      ${\cal N}= 8.02 \times 10^5$, red: ${\cal N}= 1.01\times
      10^6$.  Right panel: blue ${\cal N}= 8.31 \times 10^5$, red
      ${\cal N}=1.01\times 10^6$, cyan ${\cal N}= 1.46 \times 10^6$.
      Dotted and solid black lines have slopes $-2$ and $-8/3$,
      respectively.  }
    \label{fig_k-sharp}
\end{center} \end{figure}

Figure~\ref{fig_k-sharp}~presents the defocusing case in which a constant dissipation acts only on the  lowest few wavenumbers between $k_{\min}$ and $k_0$. Adjacent to the dissipation region, a wide plateau appears at low $k$, for which no theoretical description currently exists. The presence of this plateau provides an effective IR cut-off, limiting the total number of waves $\mN$ from above and the effective vertex $k^2/\mN$ from below. The scale at which the plateau begins was identified in \cite{Pre} as the mean distance between vortex pairs; at larger scales, the fluctuations become uncorrelated, giving rise to the plateau.  
The level $n_d$ at the dissipation scale is determined by flux balance, $\gamma n_d\simeq Q$, and may therefore be either larger or smaller than the plateau value $n_p$. Increasing $\gamma$ reduces $n_d$ and raises $n_p$, while keeping $\lambda n_dn_p$ approximately constant. This likely indicates that the interaction between these modes is essential for maintaining the flux.

 Figure~\ref{fig_k-sharp} also illustrates the nonlocality of strong turbulence in the defocusing case. In the left panel, the two spectra for $k_d=1$ exhibit a weak-turbulence regime near the pumping region, $n_k\propto k^{-2}$, which transitions into the strong-turbulence spectrum $n_k\propto k^{-8/3}$. By contrast, the spectrum with $k_d=1/2$ and $k_{\min}=1/8$ contains more long-wavelength modes but lacks the weakly turbulent part. The right panel shows that increasing dissipation --- thus suppressing a few low-$k$ modes between $k=1/8$ and $k=1/2$ --- dramatically enhances the turbulence level up to $k=16$. Even though the spectral slopes remain close to $8/3$ for high amplitudes and low $k$, the total number of waves does not scale as $\mN^{2/3}$, in contrast to the prediction of ~\eqref{minus}.

\end{document}